\documentclass[epj,final]{svjour}
\newcommand{\pdiff}[2]{\frac{\partial #1}{\partial #2}}

\newcommand{\dddot}[1]{\overset{{\,\bf...}}{#1}}

\newcommand{\rK}[1]{\left( #1 \right)}           
\newcommand{\eK}[1]{\left[ #1 \right]}           
\newcommand{\gK}[1]{\left\{ #1 \right\} }        

\newcommand{\Var}{{\rm Var}}

\usepackage{amstex}
\usepackage{epsfig}
\begin{document}
\title{Finite temperature dynamics of vortices in the two dimensional
 anisotropic Heisenberg model}
\titlerunning{Dynamics of non-planar vortices \dots}
\author{Till Kamppeter\inst{1} \and Franz G.\ Mertens\inst{1} \and
  Angel S\'anchez\inst{2} \and
  A.\ R.\ Bishop\inst{3} \and 
  Francisco Dom\'\i nguez-Adame\inst{4} \and
  N.~Gr{\o}nbech-Jensen\inst{3}}
\authorrunning{T.\ Kamppeter et al.}
\institute{Physikalisches Institut, Universit\"at Bayreuth, D-95440
  Bayreuth, Germany \and
  Grupo Interdisciplinar de Sistemas Complicados, 
  Departamento de Matem\'aticas,
  Universidad Carlos III de Madrid,\\ 
  E-28911 Legan\'es (Madrid), Spain \and
  Theoretical Division and Center for Nonlinear Studies, Los Alamos 
  National Laboratory, New Mexico 87545, U.S.A.\ \and
  Grupo Interdisciplinar de Sistemas Complicados, 
  Departamento de F\'\i sica de Materiales,
  Facultad de F\'\i sicas,\\
  Universidad Complutense, E-28040 Madrid, Spain\\ 
\email{till.kamppeter@@theo.phy.uni-bayreuth.de,\,franz.mertens@@theo.phy.uni-bayreuth.de,\,anxo@@math.uc3m.es,\\ arb@@idun.lanl.gov, adame@@valbuena.fis.ucm.es,\,\,ngj@@viking.lanl.gov}}
\date{Received: \today / Revised version: }
\abstract{We study the effects of finite temperature on the
dynamics of non-planar vortices in the classical, two-dimensional 
anisotropic Heisenberg model with XY- or easy-plane symmetry.
To this end, we analyze a generalized Landau-Lifshitz equation 
including additive white noise and Gilbert damping. 
Using a collective variable theory with no adjustable parameters we
derive an equation of motion for the vortices
with stochastic forces which are shown to
represent white noise with an effective diffusion constant
linearly dependent on temperature.  
We solve these stochastic equations of motion by means of a Green's
function formalism and obtain the mean
vortex trajectory and its variance.
We find a non-standard time dependence for the variance of the components
perpendicular to the driving force. 
We compare the analytical results with Langevin dynamics simulations 
and find a good agreement up to temperatures of the order 
of 25\% of the Kosterlitz-Thouless transition temperature. 
Finally, we discuss the reasons why our 
approach is not appropriate for higher
temperatures as well as the 
discreteness effects observed in the numerical simulations. 
} 

\maketitle

\section{Introduction}
\label{intro}
In the past two decades, solitons and other nonlinear coherent 
excitations have become a very generic and useful 
paradigm for intrinsically nonlinear phenomena in many different fields
\cite{book1,book2,book3}.
These excitations are especially important in
low dimensional systems, in terms of their relationship to key 
questions such as the existence of long
range order, the mechanisms of phase transitions or the response to
external influences \cite{alan1}. Unfortunately, for most problems of 
interest or in applications, it is not possible to develop an exact 
theory of soliton dynamics or statistical mechanics, either because 
the corresponding equation of motion is not integrable or because
perturbation terms added to it in order to account for relevant effects
destroy integrability. As a consequence, much effort 
has been devoted to develop approximate techniques allowing one to gain 
insight into soliton behavior. Among those, a very useful procedure
is that of collective variables or coordinates \cite{sirev}, which
yields very accurate results for soliton-like objects
with a well localized spatial
structure. Collective coordinate techniques have, in addition, the 
advantages of their mathematical simplicity and their 
applicability to very many perturbed soliton-bearing equations, 
including most of those which are physically relevant. 
The validity of this kind of calculations has provided grounds to 
what is nowadays called the ``particle-like picture'' of solitons: In view
of the fact
that a global coordinate, such as their center of mass, is enough to 
describe their behavior under perturbations,  it has been concluded
that solitons can be treated as point-like particles in many situations.

One important context where the above ideas are relevant is that of 
two-dimensional (2D) magnets and their collective excitations such as 
vortices or domain walls. This is a far from an academic subject: Indeed, 
in the last few years several classes of materials have been found or
fabricated for which magnetic interactions within planes of their crystalline
structure are much stronger than between these planes, and therefore the
magnetic properties are
basically 2D. Materials in these classes
include, for instance, layered magnets (such as Rb$_2$CrCl$_4$), graphite
intercalated compounds (such as CoCl$_2$), magnetic lipid layers (such as
manganese stearate), and high $T_c$ superconductors (see references
in, e.\ g., \cite{Voelkel90}).
Many of these systems can be described by 
the classical 2D anisotropic Heisenberg model with
XY- or easy-plane symmetry, given by 
\begin{equation} \label{eq:Hamiltonian}
\hspace*{-.5cm}   H=-J \sum_{\langle m,n\rangle} \eK{S_x^mS_x^n+S_y^mS_y^n+(1-\delta)S_z^mS_z^n},
\end{equation}
where the subindices $x,y$ or $z$ stand for the spin components, 
$0<\delta\leq 1$, and $\langle m,n\rangle$ labels nearest neighbors of a square
lattice. Among its excitations, specially interesting ones are vortices,
that are planar (i.\ e., with null $S_z$ components) if $\delta>0.297$
and non-planar (i.\ e., with localized $S_z$ structure) if $\delta<
0.297$ \cite{Gouvea89,nueva2}.
Such non-planar vortices are the specific object of our study as reported
in the remainder of the paper; however, the ideas we will be discussing 
are general enough to be of interest in other, related contexts where
the system behavior is governed by soliton-like structures.

The first application of a collective variable technique to the motion of 
magnetic vortices and other nonlinear magnetic excitations 
was carried out by Thiele \cite{Thiele73,Thiele74}. 
For steady state motion, when the shape $\vec{S}(\vec{r},t)$ of
the excitation in the continuum limit is rigid, he used the travelling
wave ansatz $\vec{S}(\vec{r},t)=\vec{S}(\vec{r}-\vec{X}(t))$ with
constant velocity $\dot{\vec{X}}$ (the dot stands for derivative with 
respect to time) and derived the following equation of motion, 
\begin{equation}\vec{G}_V\times\dot{\vec{X}}+\vec{F}=0,
\label{thiele}\end{equation}
where $\vec{F}$ is a
static force, due to either an external field or the interactions with
other excitations. The gyrovector $\vec{G}_V$, in 
turn, is an intrinsic quantity, produced
by the excitation itself and depending on its specific type.
$\vec{G}_V$ is perpendicular to the XY-plane; therefore, the gyrocoupling
force $\vec{G}_V\times\dot{\vec{X}}$ is formally equivalent to the Lorentz
force.
Interestingly, Thiele's 
equation, Eq.\ (\ref{thiele}), is first order, thus leading to non-Newtonian
vortex dynamics. This is somewhat unusual, as in many cases solitons are
found to behave as Newtonian point-like particles \cite{sirev}, 
obeying Newton's second
law or its relativistic generalization. We return to this point below. 

The next step beyond Thiele's approach was not taken until very recently,
when Mertens {\em et al.}\ \cite{Mertens96} developed a
generalized collective
variable theory for nonlinear coherent excitations in
classical systems with arbitrary Hamiltonians. Previously, Wysin {\em et al.}\ 
\cite{Wysin94}
had tried to generalize Thiele's equation by allowing the vortex shape 
to depend on the vortex velocity. In this way they derived a second order 
(Newtonian) equation of motion, but it was found that it did not agree with 
the simulations \cite{Mertens96,hans}. Therefore, in \cite{Mertens96}
it was proposed that the dynamics of a single
excitation is governed by a hierarchy of equations of motion for the
excitation center $\vec{X}(t)$. In addition, the Newtonian or non-Newtonian
character of the equation of motion was clarified: It was found that 
the type of the excitation
determines on which levels the hierarchy can be truncated
consistently:
So-called gyrotropic (with $|\vec{G}_V|\neq 0$) 
excitations are governed by odd-order equations
and thus do not have Newtonian dynamics.
Non-gyrotropic excitations are described by even-order
equations, i.\ e.\ by Newton's equation in the first approximation.
This is the situation for, e.\ g., domain walls.
The theory in \cite{Mertens96} was applied to non-planar (gyrotropic)
vortices of the 2D anisotropic Heisenberg model, and it was shown that
their dynamics is 
fully captured by the third-order equation, fifth-order corrections
being negligible \cite{Mertens96}.

Since the zero temperature dynamics of non-planar vortices is completely 
understood, 
in this paper we now concern ourselves with the study of non-planar vortex
dynamics in the 2D Heisenberg ferromagnet at {\em nonzero} temperatures. The 
purpose of this research is twofold: From the theoretical point of view, 
it is important to learn whether and when the vortex motion description 
in terms of an effective particle dynamics holds, and what are its main 
characteristics. In addition, the non-Newtonian character of non-planar 
vortices could be modified by temperature, or the details of the dynamics
could change as to eliminate the need to go beyond a first order equation. 
We note that if a collective coordinate theory at finite temperature 
could be worked 
out, it would provide a first step towards a statistical mechanics description
of the model behavior in terms of a vortex gas \cite{nueva1}, 
as in the case of 
one-dimensional soliton bearing systems \cite{statmech}.
On the other hand, from the experimental point of view,
insofar as the motion of vortices
has measurable consequences in inelastic
neutron scattering \cite{exp1} and nuclear magnetic resonance 
experiments \cite{exp2}, the effects of finite temperature
on vortex dynamics can have signatures in those measurements. The study we
carry out here is then necessary if there is hope to compare the 
theoretical results to actual experiments.  

The presentation of our results proceeds as follows: Section 
\ref{sec-2} contains the study of the free and the damped vortex dynamics
and the derivation of the corresponding collective coordinate theory. 
At this point the study is still deterministic, i.\ e., at zero temperature.
Section \ref{sec-3} discusses how we incorporate
the Langevin noise term to the equations of
motion. Afterwards, the collective coordinate calculation is extended to 
the resulting Langevin-Landau-Lifshitz equation, and 
the mean vortex trajectory and its variance are computed. Section 
\ref{sec-4} contains a thorough discussion of 
the comparison of the theory to the numerical 
Langevin dynamics simulation and the discussion of the main features of
the vortex motion. Finally, Section \ref{sec-5} is devoted to
the summary of our main conclusions.

\section{Zero temperature dynamics}
\label{sec-2}
%
%
%
%
%
%
Our starting point is the 
damped Landau-Lifshitz equation, which reads
\begin{equation} \label{eq:DFT}
   \frac{{\rm d}\vec{S}^m}{{\rm d}t}=
       -\vec{S}^m\times\pdiff{H}{\vec{S}^m}-
           \epsilon\vec{S}^m\times\frac{{\rm d}\vec{S}^m}{{\rm d}t},
\end{equation}
where $\vec{S}^m$ is the spin vector at lattice site $m$, $H$ is the
Hamiltonian, in our case that of the anisotropic Heisenberg 
model (\ref{eq:Hamiltonian}), and $\epsilon$
is the damping parameter. Following Refs.\ \cite{Thiele73}, \cite{Thiele74} and
\cite{Huber82} we have chosen Gilbert damping \cite{nota1},
chiefly because it is isotropic, meaning that all the spin components 
are equally damped, in contrast to the Landau-Lifshitz damping
\cite{Iida62}. 
As stated in the Introduction, 
our approach to the problem of vortex dynamics 
will be both analytical
and numerical: We first derive equations of motion for the vortex
center $\vec{X}(t)$, and 
afterwards we compare with numerical simulations for
our model, i.\ e., with results from numerical integration of
(\ref{eq:DFT}) including noise (see Section \ref{sec-3}).
The study of the deterministic 
(i.e., zero temperature) case we present in this section is an obviously
necessary first step in order to be able to understand later
the problem for the full Langevin equation.

To proceed, following \cite{Mertens96}
we assume that the shape of a collective
excitation depends on the velocity $\dot{\vec{X}}$ and, as
shown in \cite{Mertens96}, in general also on higher order derivatives
of $\vec{X}(t)$. The corresponding {\em generalized travelling wave An\-satz} is
\begin{equation} \label{eq:trav_wave}
   \vec{S}(\vec{r},t)=\vec{S}(\vec{r}-\vec{X},\dot{\vec{X}},\ddot{\vec{X}},
           \ldots,\vec{X}^{(n)})  ,
\end{equation}
which yields an $(n+1)$-th order differential equation for $\vec{X}(t)$. As
mentioned above, for
gyrotropic excitations only odd-order equations are relevant, and, 
in the case of the non-planar vortices, it turned out
that the third-order equ\-ation is sufficient to describe
accurately all simulations without damping
\cite{Mertens96}. Therefore, in this paper we use the {\em Ansatz}
(\ref{eq:trav_wave}) with $n=2$ and apply it to the general case, 
i.e., in the presence of damping.
Instead of using the Hamiltonian procedure described in \cite{Mertens96},
we will obtain the collective variable equations of motion in a much
more direct way by performing the 
following operations with (\ref{eq:DFT}): leaving out damping 
for the moment, we calculate
\begin{multline} \label{eq:wysin_eq10}
   \vec{S}\rK{\pdiff{\vec{S}}{X_i}\times\frac{{\rm d}\vec{S}}{{\rm d}t}}=
   -\vec{S}\rK{\pdiff{\vec{S}}{X_i}\times
      \eK{\vec{S}\times\frac{\delta H}{\delta \vec{S}}}}=\\
 = -S^{2}\frac{\delta H}{\delta \vec{S}}\pdiff{\vec{S}}{X_i}=
      -S^{2}\pdiff{{\cal H}}{X_i}
\end{multline}
with $i=1,2$ in the case of our 2D system. ${\cal H}$ is the
Hamiltonian density. According to our ansatz we
insert on the l.h.s.\
\begin{equation} \label{eq:dsdt}
   \frac{{\rm d}\vec{S}}{{\rm d}t}=
   \pdiff{\vec{S}}{X_j}\dot{X}_j+
      \pdiff{\vec{S}}{\dot{X}_j}\ddot{X}_j+
      \pdiff{\vec{S}}{\ddot{X}_j}\raisebox{0.3mm}{$\dddot{X}_j$}  ,
\end{equation}
integrate over $\vec{r}$ and divide by $S^2$. In this way we obtain the
same third-order equation as that obtained in Ref.\ \cite{Mertens96},
which used Hamilton equations:
\begin{equation} \label{eq:3rd-ord_eq_m}
   {\bf A}\raisebox{0.3mm}{$\dddot{\vec X}$} + {\bf M}\ddot{\vec X} + {\bf G}\dot{\vec X}
   = \vec F
\end{equation}
with force $\vec F$ given by 
\begin{equation} \label{eq:F}
   F_i = -\int\!d^2r\, \pdiff{{\cal H}}{X_i}  ,
\end{equation}
gyrotensor $\vec G$ expressed as
\begin{multline} \label{eq:Gij}
   G_{ij} =
      S^{-2}\int\!d^2r\,\vec{S}\pdiff{\vec{S}}{X_i}\times
                               \pdiff{\vec{S}}{X_j}=\\
     = \int\!d^2r\,\left\{\pdiff{\phi}{X_i}\pdiff{\psi}{X_j} -
      \pdiff{\phi}{X_j}\pdiff{\psi}{X_i} \right\}  ,
\end{multline}
mass tensor $\vec M$ with components
\begin{multline} \label{eq:Mij}
   M_{ij} = 
      S^{-2}\int\!d^2r\,\vec{S}\pdiff{\vec{S}}{X_i}\times
                               \pdiff{\vec{S}}{\dot X_j}=\\
    =  \int\!d^2r\,\left\{\pdiff{\phi}{X_i}\pdiff{\psi}{\dot X_j} -
      \pdiff{\phi}{\dot X_j}\pdiff{\psi}{X_i} \right\}  ,
\end{multline}
and third-order gyrotensor $\vec A$ given by
\begin{multline} \label{eq:Aij}
   A_{ij} = 
      S^{-2}\int\!d^2r\,\vec{S}\pdiff{\vec{S}}{X_i}\times
                               \pdiff{\vec{S}}{\ddot X_j}=\\
    =  \int\!d^2r\,\left\{\pdiff{\phi}{X_i}\pdiff{\psi}{\ddot X_j} -
      \pdiff{\phi}{\ddot X_j}\pdiff{\psi}{X_i} \right\}  . 
\end{multline}

The classical spin is constrained to have a fixed magnitude which we
set to unity. Therefore, we will 
evaluate below the expressions on the right hand sides using canonical
fields $\phi=\arctan(S_y/S_x)$ and $\psi=S_z$ for the spin vector: 
\begin{equation} \label{eq:S_m_phi}
\hspace*{-1.2cm}   \vec{S} = \sqrt{1-\psi^2}\cos\phi\,\vec{e}_x 
           + \sqrt{1-\psi^2}\sin\phi\,\vec{e}_y
           + \psi\,\vec{e}_z.
\end{equation}
At this time, we move on to 
consider the Gilbert damping term in (\ref{eq:DFT}). 
The same operation sequence as above yields:
\begin{multline} \label{eq:wysin10_w_damp}
      \epsilon\vec{S}\eK{\pdiff{\vec{S}}{X_i}\times 
         \rK{\vec{S}\times\frac{{\rm d}\vec{S}}{{\rm d}t}}}=
      \epsilon S^{2}\pdiff{\vec{S}}{X_i}\frac{{\rm d}\vec{S}}{{\rm d}t}=\\
 \!\!\!\!  =  \epsilon S^{2}\eK{
         \pdiff{\vec{S}}{X_i}\pdiff{\vec{S}}{X_j}\dot{X}_j+
         \pdiff{\vec{S}}{X_i}\pdiff{\vec{S}}{\dot{X}_j}\ddot{X}_j+
         \pdiff{\vec{S}}{X_i}\pdiff{\vec{S}}{\ddot{X}_j}\raisebox{0.3mm}{$\dddot{X}_j$}}.
\end{multline}
An integration over $\vec{r}$ gives three terms which can be combined
with the three terms on the l.h.s.\ of (\ref{eq:3rd-ord_eq_m}),
i.\ e., the damping appears in every order
\begin{multline} \label{eq:3rd-ord_damp}
   \rK{{\bf A} + {\bf a}}\!\raisebox{0.3mm}{$\dddot{\vec X}$} + 
      \rK{{\bf M} + {\bf m}}\ddot{\vec{X}} + 
      \rK{{\bf G} + {\bf g}}\dot{\vec{X}} =\\ =
   \hat{\bf A}\raisebox{0.35mm}{$\dddot{\vec X}$} + 
      \hat{\bf M}\ddot{\vec{X}} + 
      \hat{\bf G}\dot{\vec{X}} =
   \vec{F}  .
\end{multline}
The components of the damping contribution to the tensors are
\begin{multline} 
   g_{ij} =
      \epsilon\int\!d^2r\,\pdiff{\vec{S}}{X_i}
                  \pdiff{\vec{S}}{X_j} =\\
 \!\!\!\!\!  =  \epsilon\int\!d^2r\,
                  \left\{(1-\psi^2)\pdiff{\phi}{X_i}\pdiff{\phi}{X_j} +
      \frac{1}{1-\psi^2}\pdiff{\psi}{X_j}\pdiff{\psi}{X_i} \right\}
 , \label{eq:D1ij} 
\end{multline} 
\begin{multline} 
   m_{ij} = 
     \epsilon\int\!d^2r\,\pdiff{\vec{S}}{X_i}
                \pdiff{\vec{S}}{\dot X_j} =\\
 \!\!\!\!\!\!\!   = \epsilon
     \int\!d^2r\,\left\{(1-\psi^2)\pdiff{\phi}{X_i}\pdiff{\phi}{\dot X_j} +
     \frac{1}{1-\psi^2}\pdiff{\psi}{\dot X_j}\pdiff{\psi}{X_i} \right\} 
 ,  \label{eq:D2ij} 
\end{multline} 
\begin{multline} 
   a_{ij} 
    =   \epsilon\int\!d^2r\,\pdiff{\vec{S}}{X_i}
                  \pdiff{\vec{S}}{\ddot X_j} =\\
 \!\!\!\!\!\!\!     = \epsilon
       \int\!d^2r\,\left\{(1-\psi^2)\pdiff{\phi}{X_i}\pdiff{\phi}{\ddot X_j} +
       \frac{1}{1-\psi^2}\pdiff{\psi}{\ddot X_j}\pdiff{\psi}{X_i} \right\}
 . \label{eq:D3ij}
\end{multline}
We note that the first-order part of (\ref{eq:3rd-ord_damp}) was already derived
by Thiele \cite{Thiele73}.

Now, we address the problem of the 
explicit calculation of all the tensor components. This is possible only if
the {\em dynamic} structure of the collective excitation is known. The
Hamiltonian density derived from (\ref{eq:Hamiltonian}) reads \cite{Gouvea89}
\begin{multline}
   {\cal H} =
      \frac{JS^2}{2}
      \Bigg\{(1-\psi^2) (\vec{\nabla}\phi)^2 +
      \delta [4 \psi^2 - (\vec{\nabla}\psi)^2] + \\ +
      \frac{1}{1-\psi^2}{(\vec{\nabla}\psi)^2}\Bigg\}  .
   \label{eq:H_density}
\end{multline}
In \cite{Mertens96} the Hamilton equations were considered for a
non-planar vortex in the center 
of a circular system with free boundary conditions. The vortex
structure is complicated in an inner region $0 \le r \le a_c \approx
3r_{\rm v}$, where 
\begin{equation} \label{eq:r_v}
   r_{\rm v} = \frac{1}{2} \sqrt{\frac{1-\delta}{\delta}}
\end{equation}  
characterizes the vortex core \cite{Gouvea89}. $\delta$ is the
anisotropy parameter in (\ref{eq:Hamiltonian}). Recalling that 
non-planar vortices
are stable for $0<\delta<0.297$ for a square lattice.
we will use $\delta=0.1$ for our simulations.
We note that the inner region contributes very little to the integrals 
in (\ref{eq:Mij}), (\ref{eq:Aij}) and
(\ref{eq:D1ij})--(\ref{eq:D3ij});
except for (\ref{eq:Gij}), the dominant contributions stem from the
outer region $a_c \le r \le L$, if we choose a large system radius $L$.
Here
the vortex has the following dynamic structure, which is known to be
a very accurate description from simulations \cite{Mertens96}:
\begin{equation} \label{eq:dyn_phi_psi}
   \phi=\phi_0+\phi_1+\phi_2  , \quad
   \psi=\psi_0+\psi_1+\psi_2
\end{equation}
with
\begin{equation} \label{eq:static_structure}
            \phi_0 = q\tan^{-1}\frac{x_2}{x_1},
\end{equation}
\begin{equation}
      \phi_1 = p(x_1\dot{X}_1+x_2\dot{X}_2),
\end{equation}
\begin{equation}
            \phi_2 = \frac{q}{8\delta}\ln \frac{r}{eL}(x_2\ddot{X}_1-x_1\ddot{X}_2),
\end{equation}
\begin{equation}
            \psi_0 \sim p\sqrt{\frac{r_{\rm v}}{r}}\exp(-r/r_{\rm v}),
\label{jolin}
\end{equation}
\begin{equation}
             \psi_1 = \frac{q}{4\delta r^2}(x_2\dot{X}_1-x_1\dot{X}_2),
\end{equation}
and
\begin{equation}
\label{eq:static_structure2}
            \psi_2 = \frac{p}{4\delta}(x_1\ddot{X}_1+x_2\ddot{X}_2).
\end{equation}
Here $q=\pm 1$ is the vorticity and $p=\pm 1$ is the polarization, 
which determines to which side the out-of-plane structure of the 
vortex points. 
Straightforward integrations then yield the expressions of the tensor
components:
\begin{equation}
   \label{eq:G}
   G_{ij} =  G \epsilon_{ij}  , \quad G  = 2\pi pq, 
\end{equation}
\begin{equation}
   \label{eq:M}
   M_{ij} =  M \delta_{ij}  , \quad
      M  = \frac{\pi q^2}{4\delta}\ln\frac{L}{a_c} + C_{M}, 
\end{equation}
\begin{equation}
   \label{eq:A}  
   A_{ij} =  A \epsilon_{ij}  , \quad 
      A  = \frac{G}{16\delta}\rK{L^2-a_c^2} + C_{A},
\end{equation}
\begin{equation}
   \label{eq:g}
   g_{ij} =  g\delta_{ij}  , \quad
      g  = \epsilon \pi q^2 \ln\frac{L}{a_c} + C_{g},
\end{equation}
\begin{equation}
   \label{eq:a}  
   m_{ij} =  m\epsilon_{ij}  , \quad
      m  = \epsilon \frac{G}{4} \rK{L^2-a_c^2} + C_{m},
\end{equation}
and
\begin{multline}
\label{copon2}
   a_{ij} =  a\delta_{ij}, \quad
      a  = \epsilon \frac{\pi q^2}{8\delta}
                   \Bigg\{\frac{1}{2}\rK{L^2\ln L-a_c^2\ln a_c} -\\
                       \frac{1}{4}\rK{L^2-a_c^2}\Bigg\} + C_{a},
\end{multline}
where $\delta_{ij}$ is the 2D unit matrix, $\epsilon_{ij}$ is the 
antisymmetric tensor, and the different constants $C$ are the contributions
from the inner region of the vortex. 
We see that in every odd-order of (\ref{eq:3rd-ord_damp}) a symmetric
damping matrix is combined with an antisymmetric normal (non-damping)
matrix, and vice versa for the even orders. Moreover the size
dependence of the $n$-th order damping components is the same as that of
the $(n+1)$-th order normal components. The first-order damping
elements (\ref{eq:g}) were already evaluated in 
\cite{Voelkel90} and \cite{Thiele74}.

For the solution of the equation of motion (\ref{eq:3rd-ord_damp}) we
proceed as in Ref.\ \cite{Mertens96}: We
consider small displacements $\vec{x}$ from a mean trajectory
$\vec{X}^0$, on which the vortex is driven by $\vec{F}$

\begin{equation} \label{eq:mean_displacement}
   \vec X(t) = \vec X^0(t) + \vec x(t)  .
\end{equation}
We will denote the components of $\vec x$ by $x_1$ and $x_2$, with 
the {\em caveat} that they should not be confused with the components
of $\vec r$ in Eqs.\ (\ref{eq:static_structure}) through 
(\ref{eq:static_structure2}).
In view of our simulations, 
we consider the situation where the force is
always pointing in the 
$X_1$-direction and expand to first-order around $X_1(0)=R_0$ (this is
justified because in our simulations $F_0$, and even more $F_0'$, is
very small), i.\ e., 
\begin{equation}
\label{copon1}
   F = F_0 + F_0'x_1  .
\end{equation}
For $X_i^0(t)$ we obtain two coupled linear third-order equations. Taking
initial conditions $X^0_1(0)=R_0$, $X^0_2(0)=0$ the solutions are
\begin{eqnarray} \label{eq:X0_t}
   X^0_1&=&R_0+\frac{F_0}{F_0'}[\exp(t/\tau)-1]  , \\
   X^0_2&=&\frac{G}{g}\frac{F_0}{F_0'}[\exp(t/\tau)-1]  ,
\label{eq:X02_t}
\end{eqnarray}
where $\tau$ is determined by a cubic equation. The mean trajectory is
a straight line $X^0_2=G/g(X^0_1-R_0)$, 
which slightly deviates from the $X_2$-axis. The angle $g/G$ is
small because $g \sim \epsilon$, where we choose small damping
constants $\epsilon$ in the simulations. 
As $\tau$ is of the
order of $G^2/(gF_0')$ it is very large, in fact much larger than our
integration times. Therefore one can expand (\ref{eq:X0_t}) and one
get a constant velocity on the mean trajectory:
$\dot{X}^0_1=gF_0/G^2$, $\dot{X}^0_2=F_0/G$.

The motion around the mean trajectory is obtained by solving the two
coupled linear third-order equations for the 
displacements $\vec{x}(t)$ using the {\em Ansatz}
\begin{equation} \label{eq:ansatz}
   x_i=x_i^0\exp[-(\beta-i\omega)t]  .
\end{equation}
We find
\begin{multline} \label{eq:lambda12}
   \beta-i\omega=\frac{\pm i M + m}{2(A \pm i a)} \pm \\
      \pm\frac{\sqrt{(\pm i M + m)^2 - 
      4(A \pm i a)(G \pm i g)}}{2(A \pm i a)}, 
\end{multline}
with amplitude ratios $\kappa=x_2^0/x_1^0=\pm 1$ and phase differences
$\pm\pi/2$, where we have set $F_0'=0$ for simplicity. With $F_0' \neq
0$ Eq.\ (\ref{eq:lambda12}) becomes even more complicated and $|\kappa|
\neq 1$. The separation of real and imaginary parts leads to
cumbersome formulas. Therefore we compute the frequencies
$\omega_{1,2}$ and the relaxation constants $\beta_{1,2}$ as a
function of the parameters $\epsilon$ and $L$; we choose
$q=p=1$ for the charges and $\delta=0.1$ for the anisotropy.
The $a_c$-dependent parts in (\ref{eq:M})-(\ref{eq:a}) can be combined
with the constants $C_M$ etc.; the combined constants can be
neglected for large systems. As for the frequencies,
$\omega_{1,2}$ turn out to be very close to each other; hence, the
important parameters will instead be their  
mean and difference. Examples of their numerical values for a system
of radius $L=24$ are
\begin{equation} \label{eq:omega_c_m}
   \omega_c=\sqrt{\omega_1\omega_2}\approx 0.05  , \quad
   \Delta\omega=\omega_2-\omega_1 \approx 0.01,
\end{equation}
for a wide range of damping values (up to $\epsilon=0.05$), whereas 
for fixed $\epsilon$ the frequencies decrease
with $1/L$ up to rather large systems ($L=5\,000$). 
Plots of $\omega_c$ and $\Delta\omega$ as a function of $\epsilon$ 
and $L$ can be found in \cite{bad} (note, however that the caption 
under Fig.\ 1 of \cite{bad} must read 48 instead of 24).

In the simulations the purpose of the damping is to dissipate the
energy which is supplied to the system by the noise. Therefore we must
know the 
range of $\epsilon$ (for a given system size) in which the frequencies
are not influenced by the damping. As shown in \cite{bad}, 
this range is defined by the condition
\begin{equation} \label{eq:L_c}
   \epsilon L \ll 5 .
\end{equation}
The relaxation constants $\beta_{1,2}$ are nearly equal and the mean
value is $\beta_c=\epsilon/8$, and for 
the above range $\omega_c$ and $\Delta\omega$ are related to the
parameters $G$, $M$, and $A$ in a very simple way
\cite{Mertens96} 
\begin{equation} \label{eq:omega_c_m_L}
   \omega_c=\sqrt{\frac{G}{A}} \sim \frac{1}{L}  , \quad
   \Delta\omega=\frac{M}{A} \sim \frac{\ln L}{L^2}  .
\end{equation}

\begin{figure}
\hspace*{1.2cm}\epsfig{file=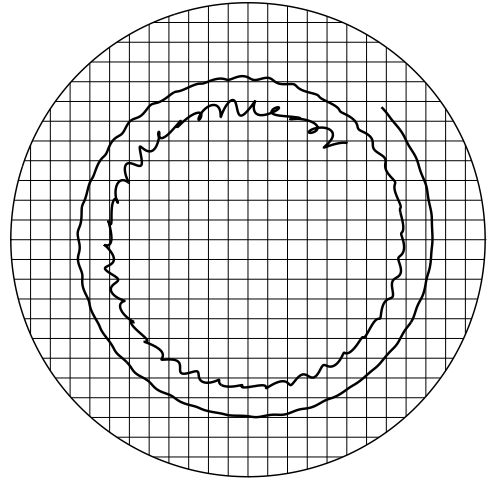, width=2.5in}
\caption{Sketch of the vortex motion as governed by the 
Landau-Lifshitz equation with Gilbert damping. The plot is approximate
and does not correspond to an actual simulation.}
\label{figure1}
\end{figure}

Finally we briefly discuss the shape of the trajectories. We first
consider the motion in a frame which is moving along the mean
trajectory $\vec{X}^0(t)$: The general solutions for the displacements
$x_i(t)$ are linear superpositions of (\ref{eq:ansatz}) with
$\omega_{1,2}$. Both $x_i(t)$ exhibit a very pronounced beat because
$\omega_{1,2}$ are nearly equal. The orbits
$x_1(x_2)$ are Lissajous curves, which can look very intricate
for certain parameter ranges. We go into the laboratory frame by
adding $\vec{X}^0(t)$. Without the splitting of $\omega_{1,2}$ we
would get a cycloid. Due to the splitting we
finally get a superposition of two cycloids, which are damped
because of $\beta_{1,2}$. A cartoon of the vortex motion on a circular
system is 
sketched in Fig.\ \ref{figure1}. Here the vortex is driven by an 
image force which points in radial direction (see Section \ref{sec-4}). 
Without damping, the mean trajectory $\vec{X}^0(t)$ would be a circle,
due to the gyrocoupling force. With damping, the circle converts to an
outward spiral. However, the radial motion is very exaggerated in the sketch,
and the same is true of the damping of the cycloidal oscillations around
the mean trajectory. The amplitude of these oscillations in fact 
remains of the order of a lattice constant for a long time. 

\section{Finite temperature dynamics}
\label{sec-3}

\subsection{Derivation of the vortex equation of motion}
\label{sec-3a}

In order to study the finite temperature dynamics of vortices, we must
introduce thermal noise in the Landau-Lifshitz equation with
damping, Eq.\ (\ref{eq:DFT}).
However, we cannot simply add independent noise terms to these three
equations, because if we do so we do not arrive to something in
the form of Langevin equations (all components of $d \vec{S}/d t$
appear in each equation due to the cross-product). 
Therefore we must first take all $d \vec{S}/d t$-terms to the l.h.s.,
casting it explicitly into a first order equation, and only then 
introduce the three white-noise terms $\eta_\alpha'(\vec{r},t)$,
yielding:
\begin{multline} \label{eq:Gilbert2}
   \frac{{\rm d}\vec{S}}{{\rm d}t}=\frac{1}{1+\epsilon^2 S^2}
       \Bigg\{-\vec{S}\times\frac{\delta H}{\delta \vec{S}}+ \\ +
           \epsilon\vec{S}\times\eK{\vec{S}\times
              \frac{\delta H}{\delta \vec{S}}}\Bigg\}
           +\boldsymbol{\eta}'
\end{multline}
with
\begin{eqnarray} \label{eq:Noise_mean}
   \langle\eta'_\alpha(\vec{r},t)\rangle &=& 0, \\
\label{eq:Noise_variance}
   \langle\eta'_\alpha(\vec{r},t)\eta'_\beta(\vec{r}\,',t')\rangle &=&
      2\epsilon k_{\rm B} T \delta(\vec{r}\,'-\vec{r}) \delta(t'-t)
      \delta_{\alpha\beta} ,
\end{eqnarray}
where $\alpha,\,\beta  =  1,\, 2,\, 3.$
Now we take $\boldsymbol{\eta}'$ to the l.h.s.\ and undo the above procedure,
i.\ e., we write (\ref{eq:Gilbert2}) in the same form as
(\ref{eq:DFT}), but with $\dot{\vec{S}}-\boldsymbol{\eta}'$ instead
of $\dot{\vec{S}}$. We thus arrive at
\begin{equation} \label{eq:Gilbert3}
   \frac{{\rm d}\vec{S}}{{\rm d}t}=
       -\vec{S}\times\frac{\delta H}{\delta \vec{S}}
          -\epsilon\vec{S}\times\frac{{\rm d}\vec{S}}{{\rm d}t}
          +\boldsymbol{\eta}
\end{equation}
with
\begin{equation} \label{eq:eta_prime}
   \boldsymbol{\eta}=\boldsymbol{\eta}'-\epsilon(\vec{S}\times\boldsymbol{\eta}')
   \enspace .
\end{equation}
If we now compute the variances of $\boldsymbol{\eta}$, we find that the width
of the distribution for the component parallel to the spin vector is
$\sigma_0=\sqrt{2 \epsilon k_{\rm B} T}$, while the widths for the
perpendicular components are $\sigma_0 \sqrt{1+\epsilon^2 S^2}$. 
In the Langevin dynamics simulation
we will apply the constraint $|\vec{S}|=1$, which means here that only
the perpendicular components are relevant. Thus we can replace
$\boldsymbol{\eta}$ in (\ref{eq:Gilbert3}) by $\boldsymbol{\eta}'$ if we correct the
widths by a factor of $\sqrt{1+\epsilon^2}$. Taking into account 
that in our simulations we will be using
values of $\epsilon$ of the order of $10^{-3}$ (see Sec.\ \ref{sec-4}),
we will neglect the
correction factor in the following.

As in the previous section, we calculate
\begin{equation} \label{eq:wysin10_w_noise}
   \vec{S}\rK{\pdiff{\vec{S}}{X_i}\times\boldsymbol{\eta}}
      =\rK{\vec{S} \times \pdiff{\vec{S}}{X_i}} \boldsymbol{\eta}
   \enspace ,
\end{equation}
integrate over $\vec{r}$ and combine this with the results of the
previous section; we have thus found the collective coordinate equation
in the presence of noise, namely
\begin{equation} \label{eq:3rd-ord_noise}
   \hat{\bf A}\raisebox{0.3mm}{$\dddot{\vec{X}}$} + 
      \hat{\bf M}\ddot{\vec{X}} + 
      \hat{\bf G}\dot{\vec{X}} =
   \vec{F} + \vec{F}^{\rm st} ,
\end{equation}
where the stochastic force is given by 
\begin{equation} \label{eq:coll_noise_1}
   F_i^{\rm st} =
   \frac{1}{S^2} \int{\rm d}^2r \rK{\vec{S} \times \pdiff{\vec{S}}{X_i}}
      \boldsymbol{\eta}(\vec{r},t).
\end{equation}
To achieve a complete understanding of the vortex dynamics as described
by Eq.\ (\ref{eq:3rd-ord_noise}), 
we need to know the mean $\langle F_i^{\rm st} \rangle$ and
the variance $\Var(F_i^{\rm st})$. We define
\begin{equation} \label{eq:FSt_i}
   F_i^{\rm st} = \int {\rm d}^2 r~ f_i^{(\alpha)} \eta_{\alpha}, \quad
   f_i^{(\alpha)} = \frac{1}{S^2} \epsilon_{\alpha\beta\gamma} S_{\beta} 
      \pdiff{S_{\gamma}}{X_i}
\end{equation}
where summation over repeated indices is implicitly understood. 
The mean is easily shown to be zero, whereas 
for the correlation functions \cite{nota2},
from Eq.\ (\ref{eq:Noise_variance}) we obtain
\begin{multline} \label{eq:Fst_var}
   \langle F_i^{\rm st}(t) F_i^{\rm st}(t') \rangle
   = \\ =2\epsilon k_{\rm B} T \delta(t-t') \int {\rm d}^2 r
      f_i^{(\alpha)}(\vec{r}) f_i^{(\alpha)}(\vec{r}) \enspace .
\end{multline}

Instead of the $S_{\alpha}$ we introduce the fields $\phi$ and $\psi$ in
(\ref{eq:S_m_phi}) and thereby fulfill the constraint
$|\vec{S}|=1$. After some algebra we obtain
\begin{multline} \label{eq:abs_Fst_var}
   \Var(F_i^{\rm st})
    = 2\epsilon k_{\rm B} T \int d^2 r 
      \Bigg\{(1-\psi^2)\rK{\pdiff{\phi}{X_i}}^2 + \\
      \frac{1}{1-\psi^2}\rK{\pdiff{\psi}{X_i}}^2\Bigg\}. 
\end{multline}
We note that in this equation
the leading contribution comes from the static vortex structure,
as given by Eqs.\ (\ref{eq:static_structure}) and (\ref{jolin}): 
\begin{multline} \label{eq:abs_Fst_var_slow}
   \Var(F_i^{\rm st})
    = 2 \pi \epsilon k_{\rm B} T \int\limits_{0}^{L} d r~  
      r \Bigg\{\frac{1-\psi_0(r)^2}{r^2} + \\
      \frac{(\psi_0'(r))^2}{1-\psi_0(r)^2}\Bigg\} .
\end{multline}
As $\psi_0$ decays exponentially, the second integral is independent
of $L$, while the first one grows logarithmically. This suggests that,
in order to approximately calculate $\Var(F_i^{\rm st})$, we can 
divide the integral
in an outer part from $a_c \leq r \leq L$ and a core part
$C(a_c)$. By doing so we can write
\begin{equation} \label{eq:Fst_var_rho}
   \Var(F_i^{\rm st})
      = 2 \epsilon k_{\rm B} T \cdot \pi \gK{\ln\frac{L}{a_c} + C(a_c)},
\end{equation}
which implies that 
the stochastic forces can be represented as white noise on
the level of the collective coordinates with the properties $\langle
F_i^{\rm st} \rangle = 0$ and 
\begin{equation} \label{eq:Fst_corr}
   \langle F_i^{\rm st}(t) F_j^{\rm st}(t') \rangle 
      = D_V \delta_{ij} \delta(t-t'),
\end{equation}
where the effective vortex 
diffusion constant $D_V$ is determined by the r.h.s.\ of
(\ref{eq:Fst_var_rho}). We recall that the diffusion constant $D$ on the 
microscopic
level, i.\ e., the one we will use in the simulations,
is $D=2 \epsilon k_{\rm B} T$. 

The core contribution $C(a_c)$ in the integral (\ref{eq:abs_Fst_var_slow})
cannot be calculated accurately by using $\psi_0(r)$ from the continuum 
limit \cite{Gouvea89}. Therefore, we have computed the full integral 
(\ref{eq:abs_Fst_var_slow}) using for $\psi_0$ an 
{\em ad-hoc} function which 
was fitted to the static vortex structure as obtained from the simulations
at zero temperature \cite{bad,Schnitzer96a}. The results for $L=24$ and 
$\delta = 0.03, 0.10$, and $0.30$ are $D_V/D = 10.02, 12.08$, and 14.18,
respectively.


\subsection{Solution of the equation of motion} 
\label{sec-3b}

We now turn to the solution of Eq.\ (\ref{eq:3rd-ord_noise}). As 
the force ${\bf F}$,
which drives the vortex, can be expanded up to first order around the
mean trajectory as discussed in Sec.\ \ref{sec-2}, we find the 
linear equation
\begin{equation}
\label{nue1}
\hat{\bf A}\raisebox{-0.1mm}{$\dddot{\vec{x}}$} +
      \hat{\bf M}\ddot{\vec{x}} +
      \hat{\bf G}\dot{\vec{x}} - {\bf f} \vec{x} = 
    \vec{F}^{\rm st} 
\end{equation}
with 
\begin{equation}
\label{nue2} 
{\bf f} = \left(\begin{array}{cc} F_0' & 0 \\ 0 & 0 \end{array}\right).
\end{equation}
Thus
we can use the Green's function formalism to obtain a formal
solution. This proceeds in two steps. First, the Green's function matrix
is obtained
from the solution to the equations
\begin{equation}
\label{Green}
\hat{\bf A}\raisebox{-0.1mm}{$\dddot{\vec{g}}_i$}+
\hat{\bf M}\ddot{\vec{g}}_i+\hat{\bf G}\dot{\vec{g}}_i -
{\bf f}\vec{g}_i= \delta(t) \vec{I}_i,
\end{equation}
where $\vec{g}_i$, 
with $i=1,2$, are the two columns of the Green's function matrix,
$\vec{I}_i$ are the corresponding columns of the identity matrix,
and suitable conditions have to be imposed on both eqs.\ 
(\ref{Green}).
Once the Green's matrix {\bf G} has been calculated, the second step is 
to solve the stochastic problem (\ref{nue1}). 
Its solution is then {\em exactly} given by
\begin{equation}
\label{soluc}
{\bf x}(t)={\bf x}_h(t)+\int_0^t\>ds\>{\bf G}(t-s)\>{\bf F}^{\rm st}(s),
\end{equation}
where ${\bf x}_h(t)$ stands for the solution of the homogeneous version
of Eq.\ (\ref{nue1}). We note that {\bf G} should not be confused with 
the gyrocoupling tensor in Eqs.\ (\ref{eq:3rd-ord_eq_m}), (\ref{eq:Gij}), 
and (\ref{eq:G}).

Let us now discuss the first part of the calculation, i.e., the computation
of the Green's matrix $\vec{G}$. Eqs.\ (\ref{Green}) above need to be suplemented
with the following conditions: {\bf G}, $\dot{\vec{G}}$, and $\ddot{\vec{G}}$ 
vanish for $t\leq 0$, and
\begin{eqnarray}
\label{cinic1} 
\vec{G}(0^+)& = &
\dot{\vec{G}}(0^+)=
\left(\begin{array}{cc} 0 & 0 \\ 0 & 0 \end{array} \right), \\
\label{cinic3} 
\ddot{\vec{G}}(0^+)&=& \hat{\bf A}^{-1} = \frac{1}{a^2+A^2}
\left(\begin{array}{cc} a & -A \\ A & a \end{array} \right).
\end{eqnarray}
In order to find the columns $\vec{g}_i$ of the Green's matrix, we take the 
{\em Ansatz}
\begin{equation}
\label{suma} 
\vec{g}_i(t)=\sum_{k=1}^6 c_k^{(i)}\left(\begin{array}{c} a_k \\ b_k 
\end{array}\right) \exp(\lambda_k t)\> \theta(t),
\end{equation}
where $\theta(t)$ is the Heaviside function, $a_k$ and $b_k$ form the 
eigenvectors belonging to the eigenvalues $\lambda_k$ of the homogeneous
problem, i.e., Eq.\ (\ref{Green}) with its r.h.s.\ set to zero,
and $c_k^{(i)}$ are the unknown amplitudes in the linear combination. The
eigenvalues are already known from the previous section: 
one is zero, one is 1$/\tau$ [see below Eq.\ (\ref{eq:X02_t})],
and the other four are 
given by Eq.\ (\ref{eq:lambda12}). All that remains is to insert the 
{\em Ansatz} (\ref{suma}) in Eq.\ (\ref{Green}) and find the 
values for $c_k^{(i)}$ from the corresponding system of algebraic 
equations. Their expression is rather cumbersome and therefore we do 
not present it here insofar as the derivation is straightforward. 

Once the $c_k^{(i)}$ and hence $\vec{G}$ are known, we can move to the
second part of the procedure, namely to find the trajectory and to 
evaluate its relevant moments. It is evident from Eq.\ (\ref{soluc})
that the mean trajectory will be exactly the same as that of the 
deterministic case, because the average of the integral of $\vec{F}^{st}$
vanishes. We will therefore concentrate on the variances,
\begin{equation}
\label{defvar} 
\sigma_{ij}^2(t)=\langle x_ix_j\rangle - \langle x_i\rangle\langle x_j
\rangle .
\end{equation} 
Using the expression (\ref{soluc}) it can immediately be seen that 
\begin{equation}
\label{vari}
\sigma_{ij}^2(t)=\sum_{k=1}^{2}\int_0^t\>dt' D_V G_{ik}(t-t')
G_{jk}(t-t'),
\end{equation}
where $G_{ij}$ stand for the elements of the Green's matrix. Once again,
the calculation is simple but tedious, due to the many terms involved 
by the product of the Green's matrix elements. Aside from this, the 
expression is easily obtained as the integrals involve only exponentials.
As an example, we present a summary of the calculation of $\sigma_{11}^2$,
which is the simplest element of the variance matrix. Nevertheless, in 
order to facilitate the presentation and the subsequent discussion we 
have made the following simplifications: i) $\omega_1=\omega_2=\omega_c$ 
and $\beta_1=\beta_2=\beta_c$, because the splittings are very small 
[see Eq.\ (\ref{eq:omega_c_m_L}) and above]; ii) $\omega_c^2+\beta_c^2
\simeq \omega_c^2$ because $\beta_c=\epsilon/8$ and $\epsilon=0.002$ in 
the simulations, implying $\beta_c$ is two orders of magnitude smaller 
than $\omega_c$ as given by Eq.\ (\ref{eq:omega_c_m}); iii) $A^2+a^2
\simeq A^2$, because $a/A={\rm O}(\epsilon)$, see Eqs.\ (\ref{eq:A}) 
and (\ref{copon2}), and iv) exponential terms involving $t/\tau$ are 
expanded to first order, because $\tau$ is much larger than our 
integration times; see below Eq.\ (\ref{eq:X02_t}). Within these 
approximations, it can be shown that 
\begin{multline}
\label{copon3}
\sigma_{11}^2(t)= \frac{D_V}{A^2\omega_c^4}\Bigg[t+\frac{1}{4\beta_c}
(1-e^{-2\beta_ct}) - \\ 
- \frac{2}{\omega_c}e^{-\beta_ct}\sin \omega_ct + \frac{1}{4\omega_c} 
e^{-2\beta_ct} \sin 2\omega_ct\Bigg].
\end{multline}
For small times, $t\ll 1/\beta_c$, we are left with an expression which
implies linear behavior plus oscillations, given by 
\begin{equation}
\label{copon4}
\sigma_{11}^2(t)= \frac{D_Vt}{A^2\omega_c^4}\Bigg(\frac{3}{2} - 
2\frac{\sin\omega_ct}{\omega_ct} + \frac{\sin 2\omega_ct}{4\omega_ct}\Bigg).
\end{equation}
We note that this function increases monotonously and that it has no 
extrema but only inflection points. For large times, $t\gg 1/\beta_c$, 
only the first term of Eq.\ (\ref{copon3}) remains, and the variance
becomes a straight line,
\begin{equation}
\label{copon5}
\sigma_{11}^2(t)= \frac{D_Vt}{A^2\omega_c^4}=\frac{D_V}{G^2}\,t, 
\end{equation}
where Eq.\ (\ref{eq:omega_c_m_L}) has been taken into account. Interestingly,
this result is identical to the one obtained by omitting the 
second- and third-order 
terms in the vortex equation of motion (\ref{eq:3rd-ord_noise}). We
thus conclude that these terms have two effects: First, they produce the 
oscillatory parts in Eq.\ (\ref{copon3}) (note that they are naturally 
connected to the cycloidal vortex trajectories), and second, for small 
times the slope of $\sigma_{11}^2$ in Eq.\ (\ref{copon4}), averaged over
the oscillations, is larger by a factor of 3/2 compared to Eq.\ (\ref{copon5}).

We do not present here the other elements of the variance matrix because
they contain even more terms than Eq.\ (\ref{copon3}). Instead, we
simply record the expression for their long time behavior, which 
is 
\begin{eqnarray}
\label{copon6} 
\sigma_{12}^2 &=& \frac{D_V}{G^2} \frac{F_0'}{2G} t^2, \\
\label{copon7} 
\sigma_{22}^2&=& \frac{D_V}{G^2}\left[t+\frac{1}{3}\left(\frac{F_0'}{G}
\right)^2 t^3\right]. 
\end{eqnarray}
The quadratic and cubic terms in $t$ appear in addition to the standard 
random walk result which is proportional to $t$. These
additional terms arise because we have allowed that the driving force 
$\vec{F}$ depends on the vortex position, see Eq.\ (\ref{copon1}). We
have considered a force in the $X_1$ direction which drives the 
vortex in the $X_2$ direction, due to the gyrocoupling force 
$\vec{G}_V\times\dot{\vec{X}}$ in Eq.\ (\ref{thiele}) or 
$\vec{G}\dot{\vec{X}}$ 
in Eq.\ (\ref{eq:3rd-ord_eq_m}), respectively. 
Therefore, only the 2-components of 
$\sigma^2$ are affected, $\sigma_{12}^2$ acquiring a factor 
$(F_0'/G)\,t$, $\sigma_{22}$ acquiring it twice.

\section{Langevin dynamics simulations}
\label{sec-4}

\subsection{Numerical procedure}
\label{subsec-41}

We begin with one vortex with its center located at a distance $R_0$ from
the middle of a circularly shaped square lattice with a
radius of $L$ lattice constants. We use free boundary conditions to get an
image antivortex which leads to a radial force on our
vortex (see \cite{Mertens96,Mertens94} and below). 
The initial spin configuration stems from an iterative program
which produces a discrete vortex structure on the lattice
\cite{Schnitzer96a}. In this way we avoid the radiation of spin waves
which would appear during the early time units if we use a continuum
approximation for the vortex shape. The parameter ranges must be chosen
very carefully for the following reasons: i) We want that
the vortex moves smoothly over the Peierls-Nabarro potential of the 
lattice; hence, the diameter $2r_v$ of the out-of-plane structure 
must be considerably larger than the lattice constant. Setting $\delta=0.1$
we find $2r_v\simeq 3$ from Eq.\ (\ref{eq:r_v}); ii) we choose a system
radius $L=24$ which provides enough space: the vortex moves outwards 
roughly on a spiral, but even for very long integration times the 
out-of-plane vortex structure should not contact the boundary, and
iii) for the same reason the initial distance $R_0$ from the middle 
of the circle should not be too large. On the other hand, $R_0$ should not
be too small; otherwise the driving force $\vec{F}$ would not be 
strong enough to overcome the pinning forces of the lattice. Letting
$R_0=10$ both conditions can be simultaneously fulfilled, if the 
damping $\epsilon$ is small enough (the larger $\epsilon$, the sooner
the vortex reaches the boundary). Note however that a small $\epsilon$
means long saturation times (see below).

For the time integration of the Landau-Lifshitz equation we use the
discrete version of (\ref{eq:Gilbert2}) where $d \vec{S}/d t$ has
already been isolated on the l.h.s.\,. In contrast to our analytical
calculations we work here with the cartesian components $S_{\alpha}$.
Therefore we explicitly take into account the constraint $\vec{S}^2=1$
by adding $\vec{S}$ times a Lagrange parameter $\lambda$ to
(\ref{eq:Gilbert2}), see Ref.\ \cite{niels}.
We form the time derivative of the constraint,
eliminate $\lambda$ and get
\begin{equation} \label{eq:Gilbert_w_constr2}
   \frac{{\rm d}}{{\rm d}t}\vec{S}=\vec{U} + \frac{\vec{S}\vec{U}}{S^2}\vec{S}
\end{equation}
with
\begin{equation} \label{eq:V}
   \vec{U} =
       \frac{1}{1+\epsilon^2 S^2}
           \rK{-\vec{S}\times\frac{\delta H}{\delta \vec{S}}+
           \epsilon\vec{S}\times\eK{\vec{S}\times
              \frac{\delta H}{\delta \vec{S}}}}
           +\boldsymbol{\eta}' \enspace ,
\end{equation}
where the site index has been omitted. We note that
(\ref{eq:Gilbert_w_constr2}) is
the same as orthogonalizing $\dot{\vec{S}}$ and $\vec{U}$ by the
Gram-Schmidt method. For the time integrations we use the same code as
in \cite{Mertens96}. In addition, the position of the vortex center,
in particular the position within a lattice cell, is determined by 
a procedure also discussed in \cite{Mertens96}.

To find a proper damping constant 
we checked the time dependence of the system energy using different damping
constants for $L=24$ and $T=0.02$. The energy at $t=0$ is 
the same as for $T=0$ and $\epsilon=0$ because the noise will be introduced 
with the first time step of the simulation. 
The energy then rises and saturates on a value
independent of $\epsilon$, but for $\epsilon > 8 \times 10^{-3}$ the
energy decreases slowly after saturation. The saturation time gets
longer with lower $\epsilon$, for $\epsilon \geq 2 \times 10^{-3}$ we
achieve acceptable saturation times $< 300$ (in units of $\hbar/(JS)$).
We have always made a pre-run of length $t_0>300$ prior to beginning
the evaluation of the simulation data. 

The difference between the energy without temperature and the saturation
energy with temperature must be the thermal energy. We computed the
mean thermal energy per spin at several temperatures and it agreed with 
$f/2 \times k_{\rm B} T$ up to $T=0.9$, $f$ being the number of degrees
of freedom per spin. For higher temperatures we
find too low values for the energy. We believe that the numerical
procedure would have to be improved if we were interested in this
regime.

\subsection{Vortex trajectory}
\label{subsec-42}

We studied the trajectory of the vortex center at different
temperatures keeping $L=24$ and $\epsilon= 2 \times 10^{-3}$ fixed. We
can distinguish three temperature regimes in which the trajectories
differ qualitatively.

\begin{figure}
\hspace*{-0.05in}\epsfig{file=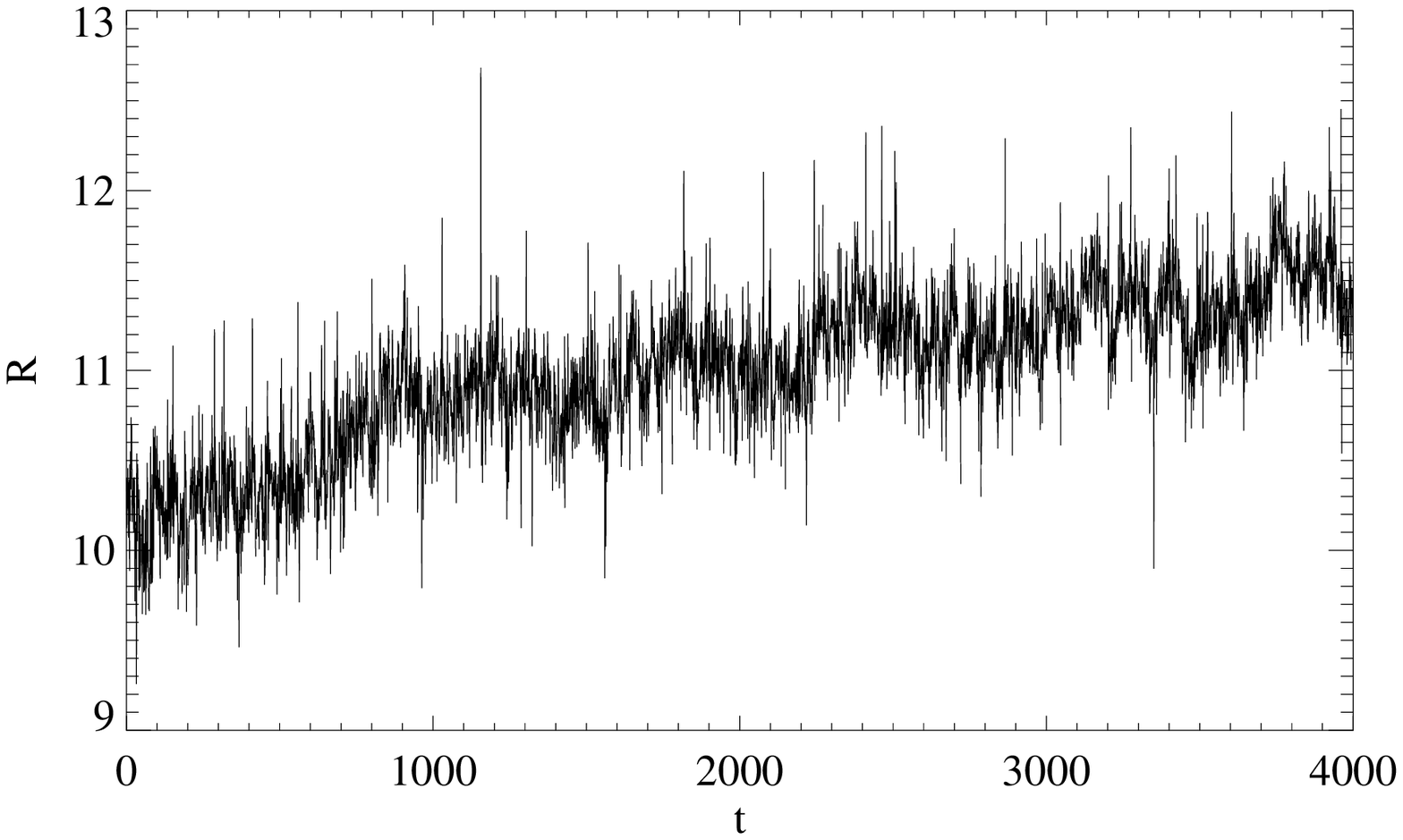, width=3.3in}

\hspace*{-0.05in}\epsfig{file=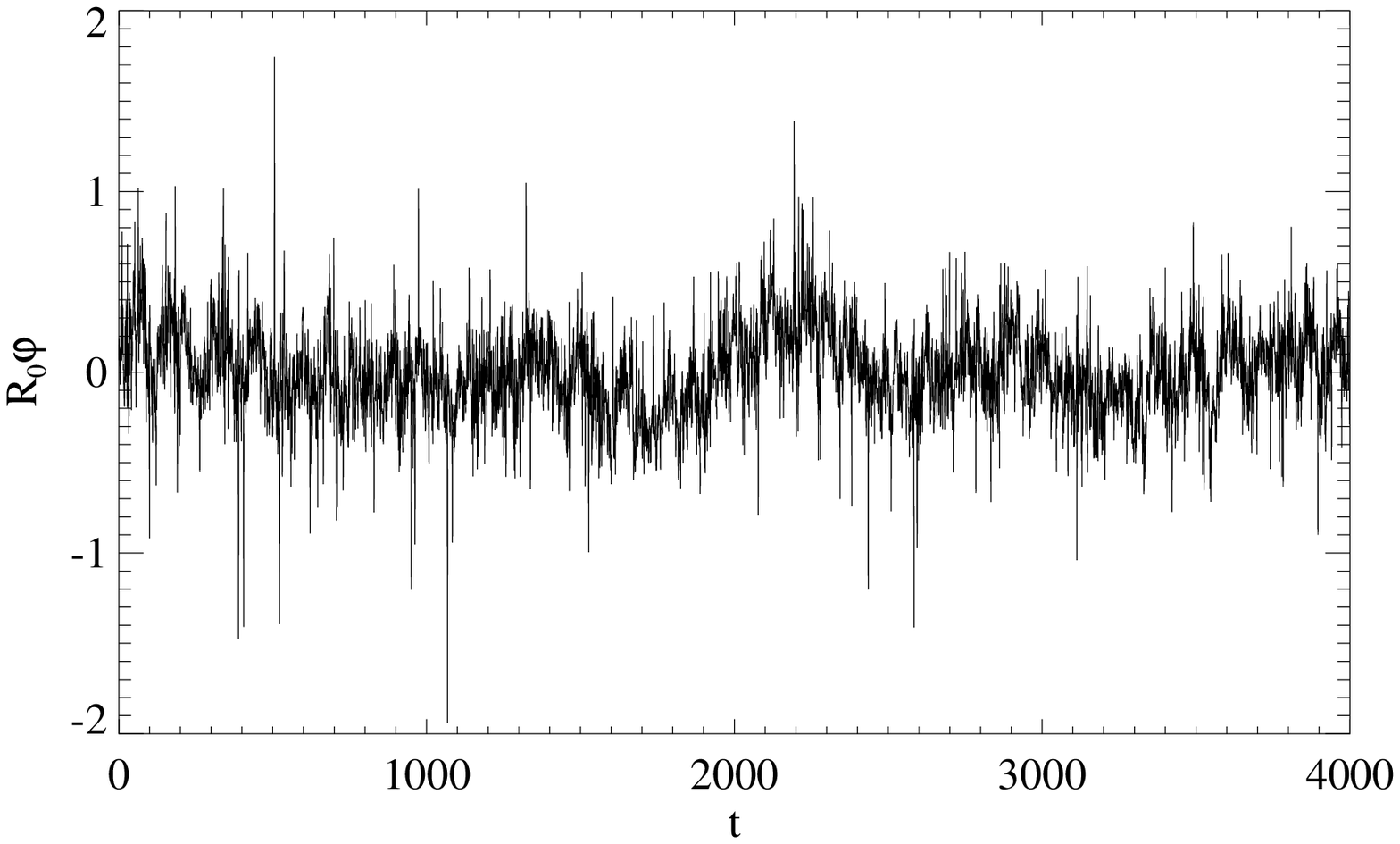, width=3.3in}

\hspace*{-0.05in}\epsfig{file=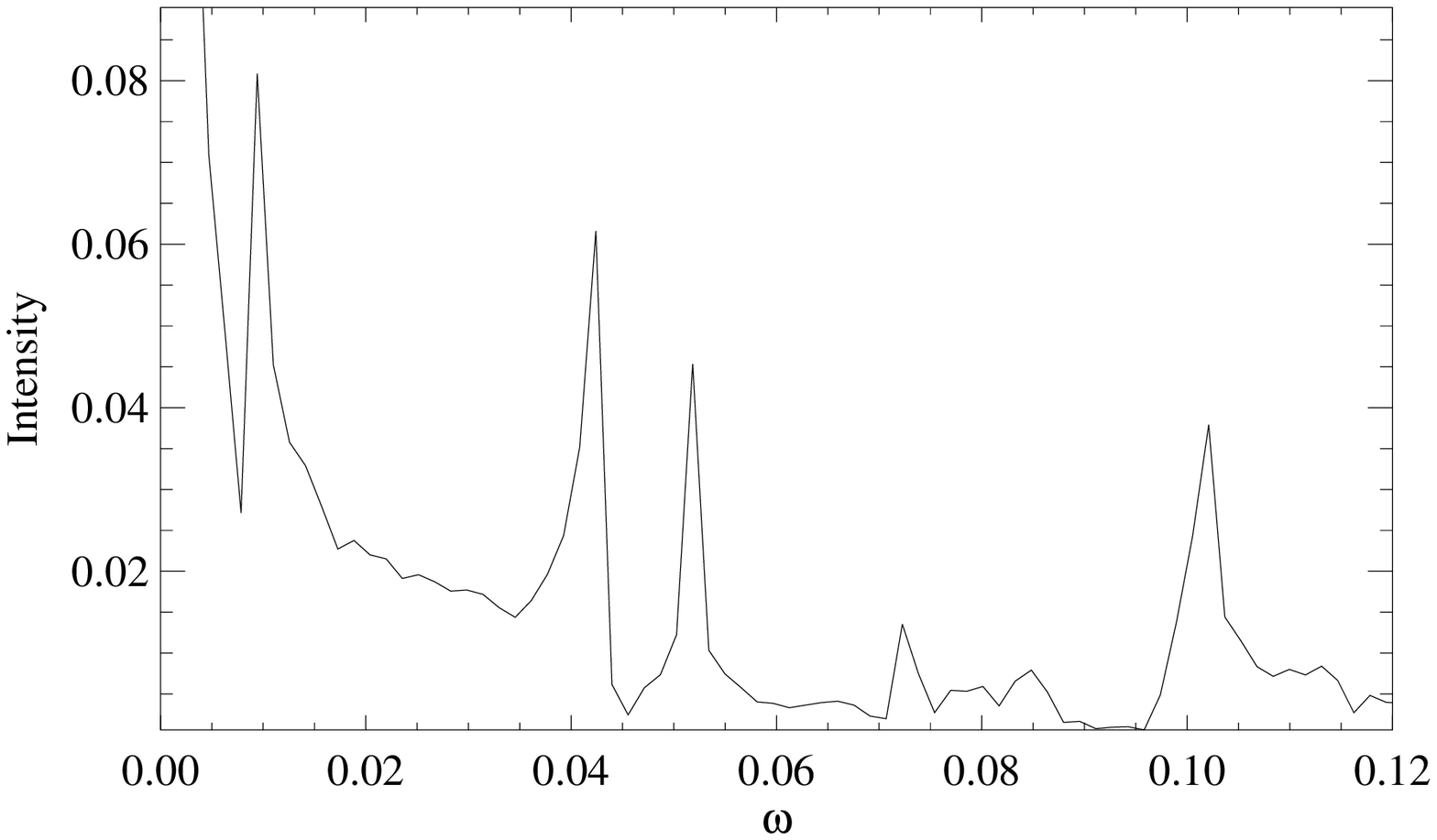, width=3.3in}
\caption{Average trajectory of a vortex for temperature $T=0.03$, 
damping $\epsilon=0.002$, system radius $L=24$ 
and an ensemble of 100 realizations. 
Upper panel: Radial coordinate of the vortex center vs time. 
Middle panel: Azimuthal displacement $\varphi=\phi(t)-\omega_0t$, where
$\omega_0$ is the angular velocity on the mean trajectory of radius 
$R_0$. Lower panel: Fourier spectrum of $R(t)$ in the upper panel. 
The spectrum of $\varphi(t)$ is very similar.} 
\label{figure2}
\end{figure}

For $0 \le T < T_3 \approx 0.05$ we observe two frequencies in the
oscillations around the mean trajectory which can be identified with
the cycloidal frequencies $\omega_{1,2}$ in (\ref{eq:lambda12}). The
intensities of the Fourier peaks at 
$\omega_{1,2}$ decrease with temperature and vanish at
$T_3$ in the background, but $\omega_{1,2}$ are constant in the whole
regime. This means that here the third order equation of motion
(\ref{eq:3rd-ord_noise}) with {\em temperature independent}
parameters can
describe the vortex dynamics. For one temperature of this regime we
plot in Fig.\ \ref{figure2} the average radial
coordinate $R(t)$ and the azimuthal displacement
$\varphi(t)=\phi(t)-\omega_0 t$. We want to stress that the plots 
present {\em averaged} results: For the computations of the vortex 
trajectories and variances we have averaged over 100 different runs
starting from the same initial configuration, which is defined as the
final configuration after a pre-run of length $t_0=1250$.
In the expression for $\varphi(t)$ above, $\omega_0=F_0/(GR_0)$ is the
frequency of the rotation on the mean trajectory which is essentially
a circle where the radius $R_0$ grows very slowly with rate
$gF_0/G^2$ due to the damping. On the mean trajectory the vortex is
driven by a radial force $F_0$ due to the image vortex at 
$R^{(\rm i)}=L^2/R_0$, which has opposite vorticity but the same
polarization \cite{Mertens94}. As the average motion is very slow
($\omega_0 \approx 2.5 \times 10^{-3}$) we can actually work in a cartesian
system and use the results
(\ref{eq:mean_displacement})-(\ref{eq:lambda12}). 
Here the $X_1$-axis
points in  
the radial direction, and the $X_2$-axis in the azimuthal
direction \cite{nueva4}. The lowest panel of 
Fig.\ \ref{figure2} shows the Fourier spectrum of
$R(t)$. In addition to $\omega_{1,2}$ one also observes the difference
$\Delta\omega=\omega_2-\omega_1$. This can be explained by working 
in polar coordinates, which is not
discussed here because the formulas become much too cumbersome.  
The peaks at higher frequencies are second harmonics of $\omega_{1,2}$.

For $T_3 < T < T_1 \approx 0.3$ we do not observe the above mentioned 
two frequencies any longer. In this regime, we found that 
some runs had to be
excluded from the average because the vortex suddenly changed its 
direction of motion. This occurs because, opposite to the case 
of the vorticity $q$, the polarization $p$ of the
vortex is not a constant of motion for a discrete system: The out-of-plane
vortex structure can flip to the other side of the lattice plane due to the
stochastic forces acting on the spins. Then $G=2\pi qp$ in Eq.\ 
(\ref{eq:G}) changes sign and thus the direction of the gyrovector
in Eq.\ (\ref{thiele}) is reversed, which implies that the direction 
of the vortex motion is reversed as well. This noise-induced switching 
between the two vortex polarizations is a very novel effect in 
itself, and hence we are developing a theory for the switching rate 
\cite{nueva3}. In this respect, it can be mentioned that switching can
also be induced by an ac magnetic field in the easy plane. As the symmetry
is broken here, such a switching occurs only for one sense of rotation,
and there is no transition back to the original state \cite{referee1}.

Finally, for $T>T_1$, a single-vortex theory as presented here
is no longer adequate because at these temperatures the probability for
the spontaneous appearance of vortex-antivortex pairs becomes too large.
These pairs can break up above the 
Kosterlitz-Thouless transition temperature $T_{KT}\simeq 0.7$ in our 
units.  Between $T_1$ and $T_{KT}$, these pairs 
interact with the initial vortex although they are not separated, thus 
introducing new forces and effects which the present theory does not 
take into account. Moreover, very recent Monte Carlo 
simulations \cite{referee2,referee3,referee4}
have revealed that for higher temperatures the vortex motion is strongly 
influenced by creation and annihilation processes: Typically, an 
unbound vortex travels only one or a few lattice spacings until it 
annihilates with the antivortex of a pair which meanwhile appeared 
spontaneously in the neighborhood. Then, the vortex of this pair 
continues the travel instead of the original vortex.

\subsection{Variances of the vortex trajectories}
\label{subsec-43}

As the vortex positions in the simulations are evaluated in polar 
coordinates, we obtain a variance matrix with elements $\sigma_{RR}^2$,
$\sigma_{R\phi}^2$ and $\sigma_{\phi\phi}^2$. Their time evolution 
\begin{figure}
\hspace*{-0.05in}\epsfig{file=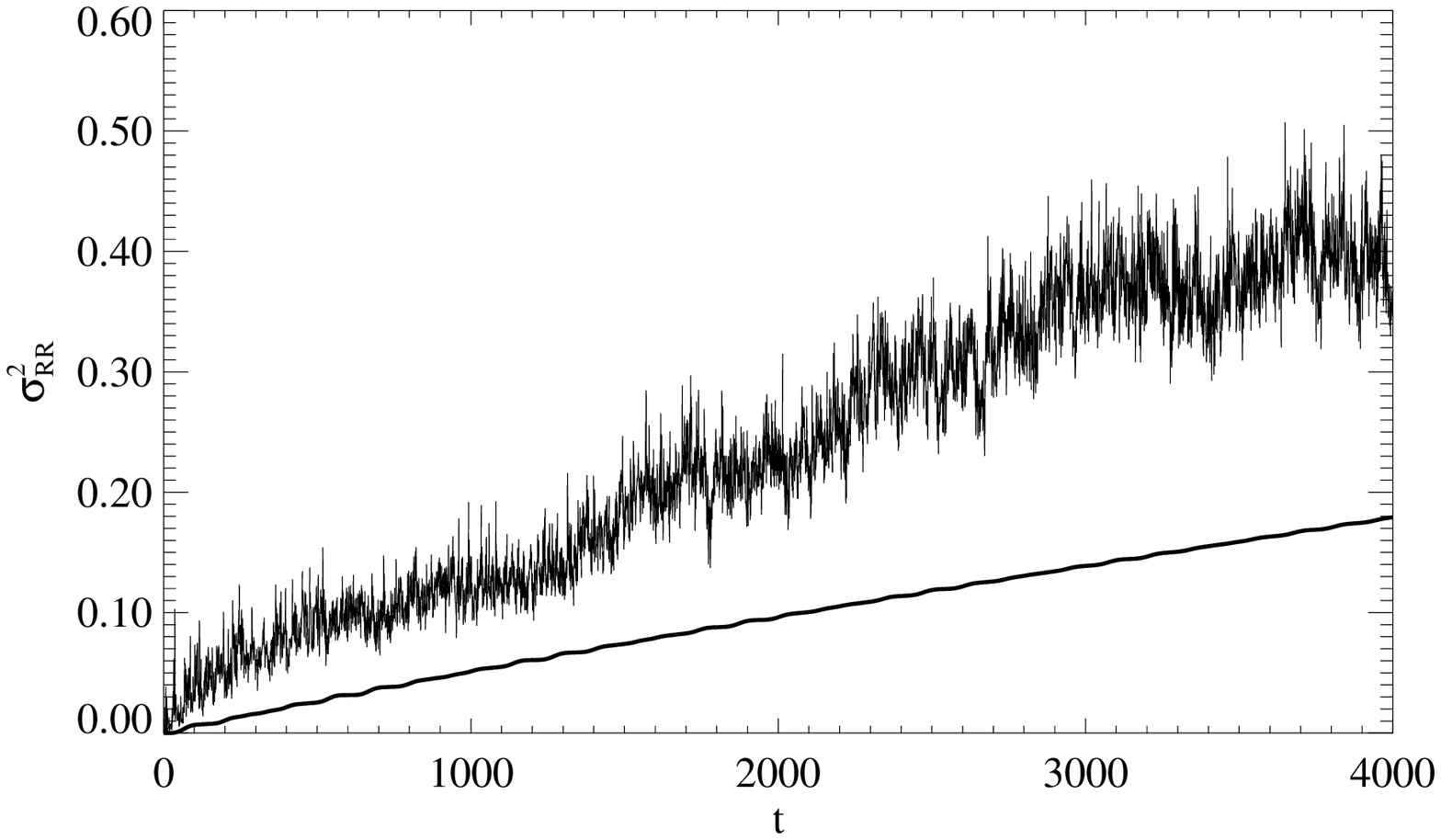, width=3.3in}

\hspace*{-0.05in}\epsfig{file=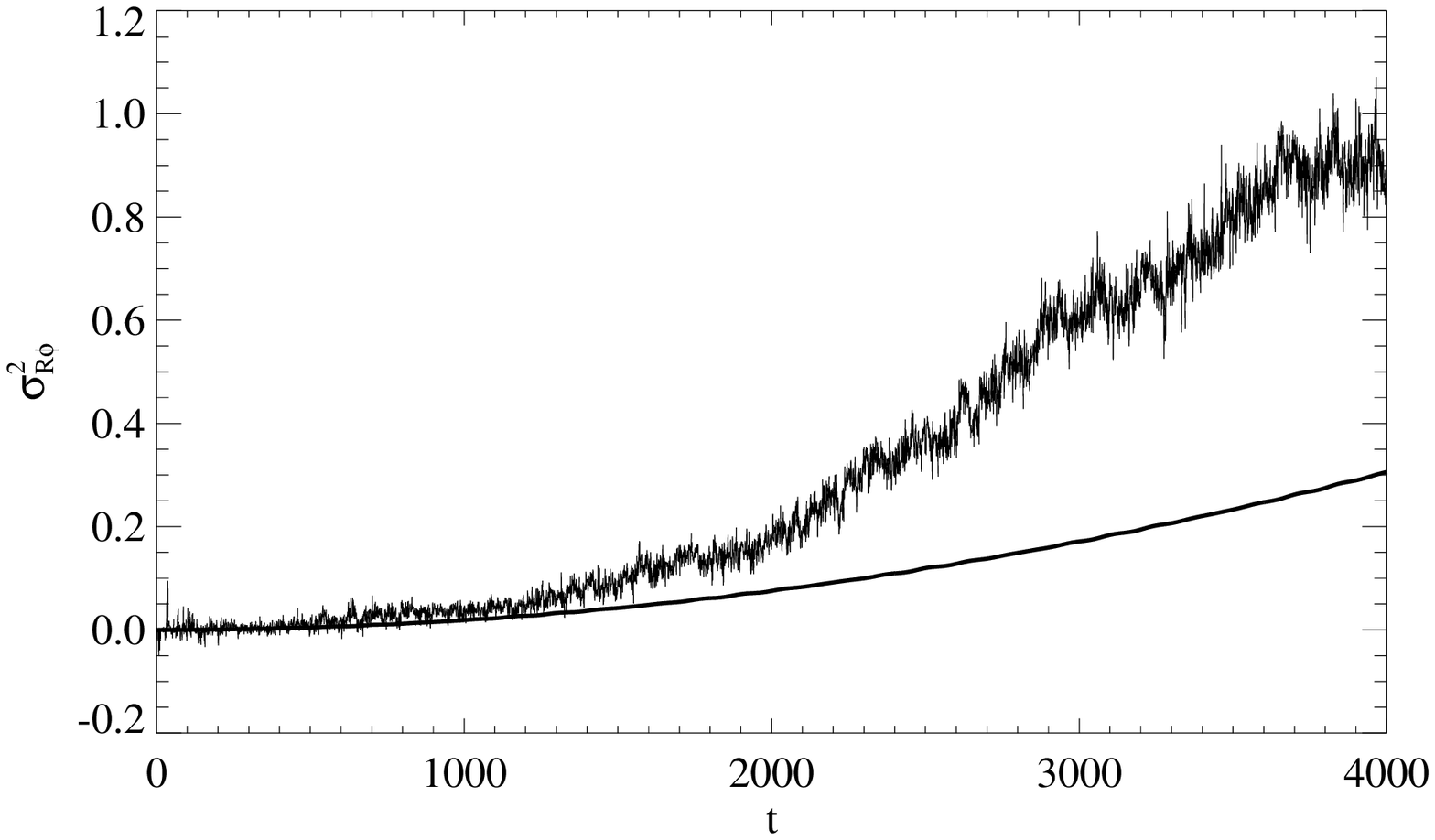, width=3.3in}

\hspace*{-0.05in}\epsfig{file=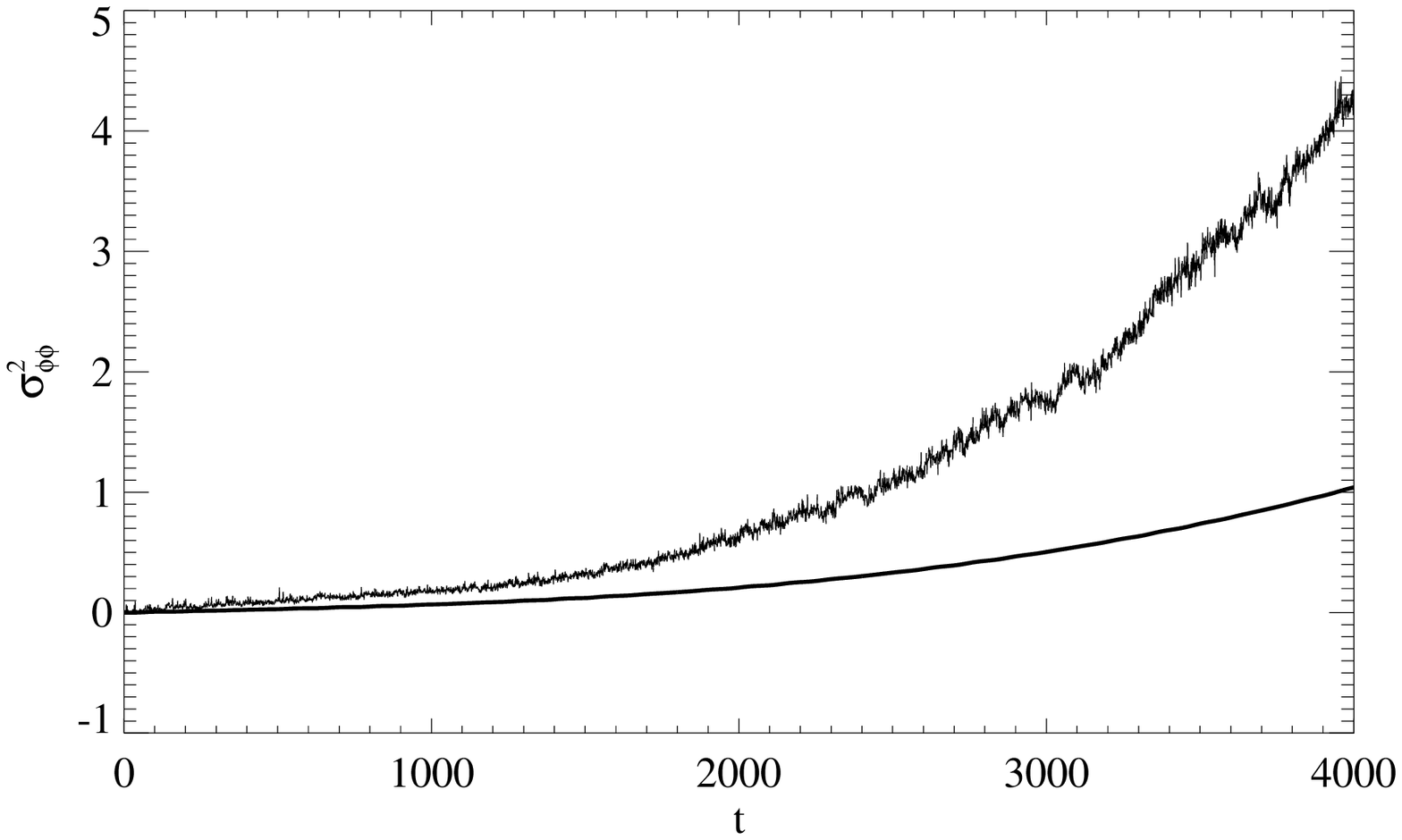, width=3.3in}
\caption{Variances of the vortex trajectory; parameters are the 
same of Fig.\ \ref{figure2}.  
From top to bottom, shown are the variance of the radial coordinate,
$\sigma_{RR}^2=\langle R^2\rangle -\langle R\rangle^2$, $\sigma_{R\phi}^2$, 
the off-diagonal elements of the variance matrix, and $\sigma_{\phi\phi}^2=
\langle (R_0\varphi)^2\rangle -\langle (R_0\varphi) \rangle^2$. In all 
three cases the lower line is the theoretical prediction without 
adjustment of parameters.}
\label{figure3}
\end{figure}
is plotted in Fig.\ \ref{figure3} for $T=0.03$, which is close to the
upper edge of the low temperature regime defined in the previous 
subsection. The solid lines are the theoretical results from Section
\ref{sec-3}, {\em without} the simplifications i)--iv) discussed 
above Eq.\ (\ref{copon3}), which were only made there to facilitate 
the discussion. As the theory has been worked out in cartesian coordinates,
the following factors appear when going over to polar coordinates: No factor
in $\sigma_{RR}^2$, a factor $\kappa=1-F_0/ (F'_0R_0)$ in 
$\sigma_{R\phi}^2$, and a factor $\kappa^2$ in front of the terms cubic 
in time in $\sigma_{\phi\phi}^2$. 

Figure \ref{figure3} shows that, for not too long times, the 
agreement between theory and simulation is astonishingly good; it is 
important to stress that no parameters were adjusted at all. Moreover, 
we worked in the continuum limit, while the simulations were
performed on a discrete system. For very long times, $t\geq 2\,000$, 
the agreement becomes poorer. This is partially due to one
simplification of the theory, namely that we have used a 
constant $R_0$ although during the simulation $R_0$ slowly increases
by several lattice constants as the trajectory is roughly a spiral 
(see the cartoon in Fig.\ \ref{figure1}). The
force term $F'_0=F'(R_0)$ increases as well, because the force increases 
when the distance to the image vortex becomes smaller. As $F'_0$ 
appears in $\sigma_{R\phi}^2$ and $(F'_0)^2$ arises in front of the
cubic term in $\sigma_{\phi\phi}^2$ in Eqs.\ (\ref{copon6}) and 
(\ref{copon7}), including this effect would lead to an improvement
of the agreement between theory and simulations. 

Aside from those discussed above, there is 
another possible reason for the discrepancy between theory and 
simulation whose consideration, unfortunately, would lead to very 
involved calculations: The integral (\ref{eq:abs_Fst_var}) for 
$D_V$ [as well as the integrals in Sec.\ \ref{sec-2} except 
(\ref{eq:Gij})] have been evaluated by placing the vortex into the
middle of the circular system. However, in the simulations the 
distance from the lattice center is $R_0$, which moreover increases
slowly. We have estimated the above integrals by expanding in $R_0/L$,
which shows that the first order terms vanish. Nevertheless, the 
second order terms yield corrections which are already of the order of 
20\% for $R_0=10$, becoming larger as $R_0$ increases. Even more,
the variance (\ref{eq:Fst_var_rho}) of the stochastic forces is actually a 
diagonal {\em tensor}, see Eq.\ (\ref{eq:Fst_var}) and \cite{bad}. Therefore, 
we get a radial diffusion constant $D^{(R)}_V$ which differs from 
the azimuthal constant $D^{(\phi)}_V$ when the vortex is not at the
center. This splitting is also of order $(R_0/L)^2$. As $D^{(\phi)}_V$
appears, e.\ g., in front of the cubic term in $\sigma_{\phi\phi}^2$, 
whereas the linear term contains $D^{(R)}_V$, it is quite possible 
that the agreement with the simulations could be improved by taking
into account the splitting of the diagonal elements of the diffusion 
tensor.

We numerically integrated up to times $t=4\,000$ (let us point out in 
this regard that this takes three weeks CPU time on a CRAY-YMP/EL for 
averages over 100 runs) because this is the characteristic 
time given by $1/\beta_c=8/\epsilon$ for the damping in the trajectories.
We should see then that the slope of the time-averaged function 
$\sigma_{RR}^2$ gradually decreases, eventually by a factor one third 
for $t\gg 1/\beta_c$ [cf.\ the discussion of Eq.\ (\ref{copon3})]. We
checked this for the theoretical results in Fig.\ \ref{figure3}, and 
found that in the simulation data this effect can be observed only 
qualitatively. For $\sigma_{R\phi}^2$ and 
$\sigma_{\phi\phi}^2$ the effect is hidden by the quadratic and cubic
terms. 

\begin{figure}
\hspace*{-0.05in}\epsfig{file=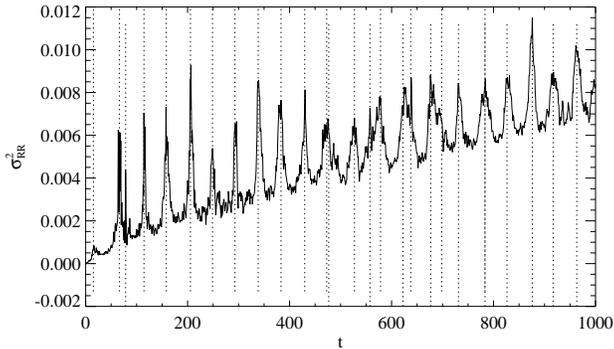, width=3.3in}
\caption{Variance of the radial vortex coordinate averaged over 1\,500 
realizations, for $T=0.003$, $\epsilon=0.002$ and $L=24$. The dashed 
vertical lines indicate the times at which the vortex center moves 
over ridges of the Peierls-Nabarro periodic potential.}
\label{figure4}
\end{figure}

We would like to stress that the strong fluctuations in Fig.\ \ref{figure3} 
(which seem to be smaller in the two lower plots because of the different
scales) arise not only due to the noise but also from discreteness effects.
This is demonstrated very clearly by the simulation in Fig.\ \ref{figure4} 
for a very low temperature ($T=0.003$) using 1\,500 realizations. We have
identified the sharp spikes as discreteness effects by comparing with 
the times when the vortex center moves over the ridges along the lattice
lines (these times are indicated as dashed vertical lines). The vortex 
energy is highest when the center is at a lattice point, and lowest in 
the middle of a cell. 

\begin{figure}
\hspace*{-0.05in}\epsfig{file=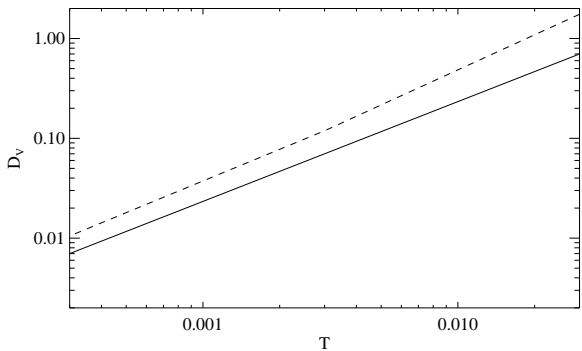, width=3.3in}
\caption{Vortex diffusion constant $D_V$ as a function of temperature,
for $\epsilon=0.002$ and $L=24$. Solid line: Theoretical results from 
Eq.\ (\ref{eq:abs_Fst_var_slow}); dashed line: Adjusted $D_V$ from
fitting the theoretical curves for $\sigma^2(t)$ to the simulation
data.}
\label{figure5}
\end{figure}

Last, but not least, we discuss the temperature dependence of the vortex 
diffusion constant $D_V$. A linear dependence is predicted by Eqs.\ 
(\ref{eq:abs_Fst_var_slow}) and (\ref{eq:Fst_var_rho}). For comparison 
with the simulations we have fitted the theoretical curves to the 
observed variances by adjusting $D_V$, which appears as a factor in 
front of all the components of $\sigma^2$. This was done for two 
temperature decades. Fig.\ \ref{figure5} shows a nearly linear 
dependence, and therefore the only difference between $D_V$ from the
simulations and the theoretical $D_V$ is a constant factor of about 
1.8 for the whole temperature regime. 

\section{Conclusions}
\label{sec-5}

In this paper, we have reported our analytical and numerical work 
regarding the effects of temperature on the dynamics of non-planar 
vortices in 2D, classical, anisotropic Heisenberg ferromagnets. 
As a preliminary result, we have described the zero 
temperature dynamics of vortices in the presence of Gilbert damping. 
We found that damping contributes to all the terms of the third order 
equation of motion for the vortex position, but its contribution 
is always an order smaller in the system size than the corresponding 
free propagation part. We have solved the equations of motion and 
qualitatively discussed the motion of the vortex, which consists of 
a mean straight trajectory plus (damped) additional oscillations. 
By means of the same analytical approach, we 
have been able to derive a third order {\em stochastic} equation of motion 
for the vortex center when thermal noise is added to the system. The 
equation shows that the effective stochastic force acting on the 
vortex is also a Gaussian white noise, whose variance depends linearly
on the temperature. We have exactly solved the stochastic equation of 
motion and obtained analytical expressions for the mean vortex 
trajectory and its variance. The variance along the coordinate perpendicular
to the direction of motion of the vortex is diffusive, i.e., it increases
linearly with time; however, other components of the variance matrix 
(the parallel-perpendicular and the parallel-parallel terms)
turn out to include nonlinear contributions coming from the fact 
that the vortex motion is perpendicular to the driving force, due to a 
Lorentz-like gyrocoupling force. 

The above summarized analytical results, obtained in the continuum limit 
of the Landau-Lifshitz equations governing the model dynamics,
have been compared to Lan\-ge\-vin dynamic simulations of the discrete 2D 
Heisenberg model. The numerical results allow us to establish three 
different temperature regimes for the vortex propagation: a low temperature
one, where the vortex motion follows essentially the third order equation 
of motion with parameters independent of temperature; a middle temperature 
one, at which traces of the oscillations arising from the third order 
equation are lost, and a high temperature regime, which is not describable
by a one-vortex approach because too many vortex-antivortex pairs arise 
in the system. Our analytical results are seen to be a good description 
of the vortex dynamics up to temperatures of the order of 10\% of the 
Kosterlitz-Thouless transition temperature. Remarkably, the analytical 
predictions, which include no adjustable parameters, agree qualitatively
well with the numerical simulations, and even quantitatively at early times. 
The agreement becomes worse for longer times due to the approximations 
involved in our theory: The calculations were made for a constant
radius of the trajectory and a constant force {\bf F} gradient,
aside from simplifications
necessary to calculate the integrals which give the parameters for the 
equation of motion. In addition, we have been able to clearly identify the
influence of discreteness in the numerical results, which cannot be captured
by our continuum theory. Finally, we have also verified that the vortex 
diffusion constant depends linearly on temperature as predicted, although
the quantitative comparison is not correct by a factor two. We thus 
conclude that the collective coordinate theory we have derived for 
vortex dynamics is a good description of the phenomena observed 
numerically at low 
and intermediate temperatures. The discrepancies between theory and 
simulations have been understood in terms of the unavoidable 
approximations involved in the calculations. Finally, we note that
for vortices quantum effects are possibly more important than for 
kinks in one-dimensional spin models, where at least a part of these
effects can be taken into account by a renormalization of the kink 
parameters. For 2D spin models, it is not clear how a {\em quantum 
vortex} should be defined. In any case, a finite lifetime and other 
novel features seem to appear \cite{referee5}. 
 
\begin{acknowledgement}
We thank Esteban Moro and Grant Lythe for discussions.
Tra\-vel between Bayreuth and Madrid is supported by ``Acciones
Integradas Hispano-Alemanas'', a joint program of DAAD (Az.\ 314-AI)
and DGES.
Travel between Europe and Los Alamos is supported by NATO
grant CRG 971090.
Work at Madrid and Legan\'es is supported by CICyT (Spain) 
grant MAT95-0325 and by DGES (Spain) grant PB96-0119.
Work at Los Alamos is supported by the United States Department of Energy.
\end{acknowledgement}


\begin{thebibliography}{99}
\bibitem{book1} {\em Nonlinearity in Condensed Matter}, edited by A.\ R.\ 
Bishop, R.\ Ecke, and J.\ Gubernatis (Springer, Berlin, 1993).
\bibitem{book2} {\em Nonlinear Coherent Structures in Physics and Biology},
edited by K.\ H.\ Spatschek and F.\ G.\ Mertens (Plenum, New York, 1994).
\bibitem{book3} {\em Fluctuation Phenomena: Disorder and Nonlinearity},
edited by A.\ R.\ Bishop, S.\ Jim\'enez, and L.\ V\'azquez
(World Scientific, Singapore, 1995). 
\bibitem{alan1} See, e.g., A.\ R.\ Bishop, in Ref.\ \cite{book3}
\bibitem{sirev} A.\ S\'anchez and A.\ R.\ Bishop, SIAM Review in press (1998).
\bibitem{Voelkel90}
   A.\,R.\,V\"olkel, F.\,G.\,Mertens, A.\,R.\,Bishop and G.\,M.\,Wysin,
   Phys.\ Rev.\ {\bf B43}, 5992 (1991).
\bibitem{Gouvea89}
   M.\,E.\,Gouvea, G.\,M.\,Wysin, A.\,R.\,Bishop and F.\,G.\,Mer\-tens,
   Phys.\ Rev.\ {\bf B39}, 11840 (1989).
\bibitem{nueva2} G.\ M.\ Wysin, Phys.\ Lett.\ A, in press.
\bibitem{Thiele73}
   A.\,A.\,Thiele,
   Phys.\ Rev.\ Lett.\ {\bf 30}, 230 (1973).
\bibitem{Thiele74}
   A.\,A.\,Thiele,
   J.\ Appl.\ Phys.\ {\bf 45}, 377 (1974).
\bibitem{Mertens96}
   F.\,G.\,Mertens, H.-J.\,Schnitzer and A.\,R.\,Bishop,
   Phys.\ Rev.~B {\bf 56}, 2510 (1997).
\bibitem{Wysin94}
   G.\,M.\,Wysin, F.\,G.\,Mertens, A.\,R.\,V\"olkel and A.\,R.\,Bishop,
   in \cite{book2}.
\bibitem{hans} A.\ R.\ V\"olkel, G.\ M.\ Wysin, F.\,G.\,Mertens, 
A.\,R.\,Bishop, and H.\ J.\ Schnitzer, Phys.\ Rev.\ B {\bf 50}, 12\,711 (1994).
\bibitem{nueva1} F.\ G.\ Mertens, A.\ R.\ Bishop, G.\ 
M.\ Wysin, and C.\ Kawabata, Phys.\ Rev.\ B {\bf 39}, 591 (1989).
\bibitem{statmech} 
   J.\ F.\ Currie, J.\ A.\ Krumhansl, A.\ R.\ Bishop and S.\ E.\ Trullinger,
Phys.\ Rev.\ B {\bf 22}, 477 (1980).
\bibitem{exp1} 
K. Hirakawa, H. Yoshizawa, J. D. Axe, and G. Shirane, Suppl. J. Phys. Soc. 
Jpn. {\bf52}, 19 (1983);
L. P. Regnault, J. P. Boucher, J. Rossat-Mignod, J. Bouillot, R. Pynn, J. Y. 
Henry, and J. P. Renard, Physica B+C {\bf 136B}, 329 (1986);
M. T. Hutchings, P. Day, E. Janke, and R. Pynn, J. Magn. Magn. Mater. {\bf 
54-57}, 673 (1986);
S. T. Bramwell, M. T. Hutchings
 J. Norman, R. Pynn, and P. Day, J. de Phys. {\bf 49}, C8-1435 (1988);
D. G. Wiesler, H. Zabel, and S. M. Shapiro, Physica B {\bf 156-7}, 292 (1989);
D. G. Wiesler, H. Zabel, and S. M. Shapiro, Z. Physik B {\bf 93}, 277 (1994);
L. P. Regnault, 
C. Lartigue, J. F. Legrand, B. Farago, J. Rossat-Mignod,
and J. Y. Henry, Physica B {\bf 156-7}, 298 (1989).
\bibitem{exp2} 
P. Gaveau, J. P. Boucher, L. P. Regnault, and Y. Henry, J. Appl. Phys. {\bf 
69}, 6228 (1991).
\bibitem{Huber82}
   D.\,L.\,Huber,
   Phys.\ Rev.\ {\bf B26}, 3758 (1982).
\bibitem{nota1} The sign of our
  damping term differs from \cite{Thiele73,Thiele74,Iida62} because we
  work with spins while these authors deal with magnetisations.
\bibitem{Iida62}
   S.\,Iida,
   J.\ Phys.\ Chem.\ Solids {\bf 24}, 625 (1963).
\bibitem{bad} T.\ Kamppeter, F.\ G.\ Mertens, A.\ S\'anchez, N.\ 
Gr\o nbech-Jensen, A.\ R.\ Bishop, and F.\ Dom\'\i nguez-Adame, in 
``Theory of Spin Lattices and Lattice Gauge
Models,'' eds.\ M.\ L.\ Ristig, K.\ A.\ Gernoth and J.\ W.\ Clark.
Springer-Verlag, Berlin (1997).
\bibitem{nota2} The correlation between
  different components of ${\bf F}^{\rm st}$ can also be calculated
  and turns out to be zero. 
\bibitem{Miyashita78}
   S.\,Miyashita, H.\,Nishimori, A.\,Kuroda and M.\,Suzuki, 
   Progress of Theor.\ Phys.\ {\bf 60}, 1669 (1978).
\bibitem{Schnitzer96a}
   H.-J.\,Schnitzer,
   {\it Zur Dynamik kollektiver Anregungen in Hamiltonschen Systemen,}
   Ph.D.-thesis, University of Bayreuth (1996).
\bibitem{niels} N.\ Gr\o nbech-Jensen and S.\ Doniach, J.\ Comp.\ Chem.\ 
{\bf 15}, 997 (1994).
\bibitem{Mertens94}
   F.\,G.\,Mertens, G.\,M.\,Wysin, A.\,R.\,V\"olkel and A.\,R.\,Bishop,
   in \cite{book2}.
\bibitem{nueva4} 
Strictly speaking, Eq.\
(\ref{eq:3rd-ord_noise}) must be solved in polar coordinates which leads to
shifts of $\omega_{1,2}$ by $\pm\omega_0$, respectively \cite{Mertens96}.
The different signs appear due to the phase differences $\pm\pi/2$, 
below Eq.\ (\ref{eq:lambda12}).
\bibitem{nueva3} Yu.\ Gaididei, T.\ Kamppeter, F.\ G.\ Mertens and 
A.\ R.\ Bishop, submitted (August 1998).
\bibitem{referee1} Yu.\ Gaididei, T.\ Kamppeter, F.\ G.\ Mertens and 
A.\ R.\ Bishop, in preparation.
\bibitem{referee2} D.\ A.\ Dimitrov and G.\ M.\ Wysin, Phys.\ Rev.\ B {\bf 
53}, 8539 (1996).
\bibitem{referee3}  J.\ E.\ R.\ Costa, B.\ V.\ Costa, and D.\ P.\ Landau, 
Phys.\ Rev.\ B {\bf 57}, 11\,510 (1998). 
\bibitem{referee4} D.\ A.\ Dimitrov and G.\ M.\ Wysin, preprint (March 1998).
\bibitem{referee5} J.\ Schliemann, F.\ G.\ Mertens, and H.\ Frahn, in 
preparation. 
\end{thebibliography}
\end{document}